# A New Method for Sensorless Estimation of the Speed and Position in Brushed DC Motors Using Support Vector Machines

Ernesto Vázquez-Sánchez, Jaime Gómez-Gil, José Carlos Gamazo-Real, and José Fernando Díez-Higuera

***Abstract*—Currently, for many applications, it is necessary to know the speed and position of motors. This can be achieved using mechanical sensors coupled to the motor shaft or using sensorless techniques. The sensorless techniques in brushed dc motors can be classified into two types: 1) techniques based on the dynamic brushed dc motor model and 2) techniques based on the ripple component of the current. This paper presents a new method, based on the ripple component, for speed and position estimation in brushed dc motors, using support vector machines. The proposed method only measures the current and detects the pulses in this signal. The motor speed is estimated by using the inverse distance between the detected pulses, and the position is estimated by counting all detected pulses. The ability to detect ghost pulses and to discard false pulses is the main advantage of this method over other sensorless methods. The performed tests on two fractional horsepower brushed dc motors indicate that the method works correctly in a wide range of speeds and situations, in which the speed is constant or varies dynamically.**

## I. Introduction

SENSORLESS techniques estimate the speed and position of motors without mechanical sensors coupled to the motor shaft, measuring only the current and/or the voltage of the motors. Sensorless techniques are not a recent idea, as is evidenced by the work of Allured and Strzelewigz [1]. Nevertheless, due to the complexity of these methods, they have not yet replaced conventional sensors such as encoders, potentiometers, tachometers, Hall effect sensors, or other mechanical sensors coupled to the motor shaft. The main advantages of these, compared to conventional sensors, are as follows: 1) decreased maintenance, number of connections, and cost of the final system and 2) an easier miniaturization process. In addition, mechanical elements are not coupled to the motor shaft in sensorless techniques. Sensorless techniques are particularly useful in fractional horsepower applications because they usually require a low cost and a low hardware complexity. Sensorless techniques function by monitoring the voltage and/or current of the motor to estimate the speed and position. The problem with the implementation of sensorless techniques is software complexity, since the models used and the noise in the current and voltage make it difficult to estimate the velocity and position of the motor [2].

Sensorless techniques in brushed dc motors can be divided into two groups: 1) those based on the dynamic dc motor model and 2) those based on the ripple component of the motor current [3]. The first group is mainly employed to estimate the speed using the dynamic model of the brushed dc motor [4]–[9]. The dynamic model uses different parameters of the brushed dc motor such as resistance, inductance, and constant electromotive force (EMF). The problem with using the parameters of the brushed dc motor is that they depend on the operating conditions, which are changing and introduce uncertainty into the speed measurement. Although these parameters can be estimated dynamically [10], [11], this solution usually leads to a nonlinear model that increases the computational cost.

Sensorless techniques based on the ripple component only monitor the brushed dc motor current and they estimate speed and position with instantaneous variations of the current [12]–[16]. The current of a brushed dc motor, shown in Fig. 1, is mainly composed of two components: the dc component and the ripple component. The dc component is responsible for providing power to the brushed dc motor. The ripple component is an alternating component and is the direct result of two effects. The first effect is the nonideal rectification that occurs in the complex brush-commutator system that connects the rotor with the external circuit. The second effect appears in the coil of the motor and is an EMF induction, with an approximately sinusoidal shape not rectified ideally by the mechanical switching system.

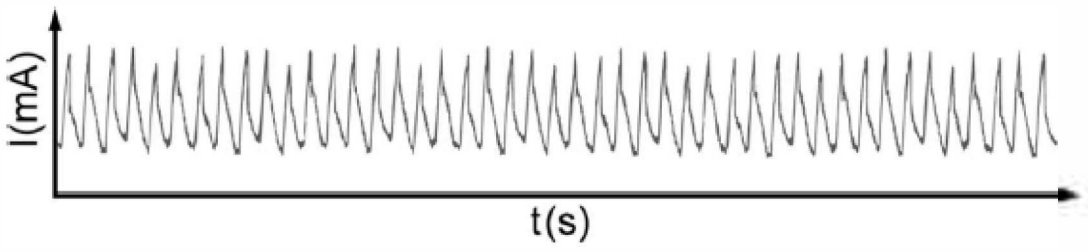

Fig. 1. Current of a brushed dc motor.

Manuscript received June 1, 2010; revised September 24, 2010, February 18, 2011, March 28, 2011, and May 27, 2011; accepted June 18, 2011. Date of publication July 14, 2011; date of current version October 25, 2011. This work was supported by the Regional 2010 Research Project Plan of Junta de Castilla y León (Spain), under VA034A10-2 Project. The work of E. Vazquez-Sanchez was supported by a grant from the *Contratación de Personal Investigador de Reciente Titulación* program, which was financed by Consejería de Educación of Junta de Castilla y León (Spain) and cofinanced by the European Social Fund.

E. Vázquez-Sánchez, J. Gómez-Gil, and J. F. Díez-Higuera are with the Department of Signal Theory, Communications and Telematics Engineering, University of Valladolid, 47011 Valladolid, Spain (e-mail: ernesto.vazquez@uva.es; rnstvaz@gmail.com; jgomez@tel.uva.es; josdie@tel.uva.es).

J. C. Gamazo-Real is with the Test Facilities Department, EADS-CASA Cassidian, 28906 Madrid, Spain.

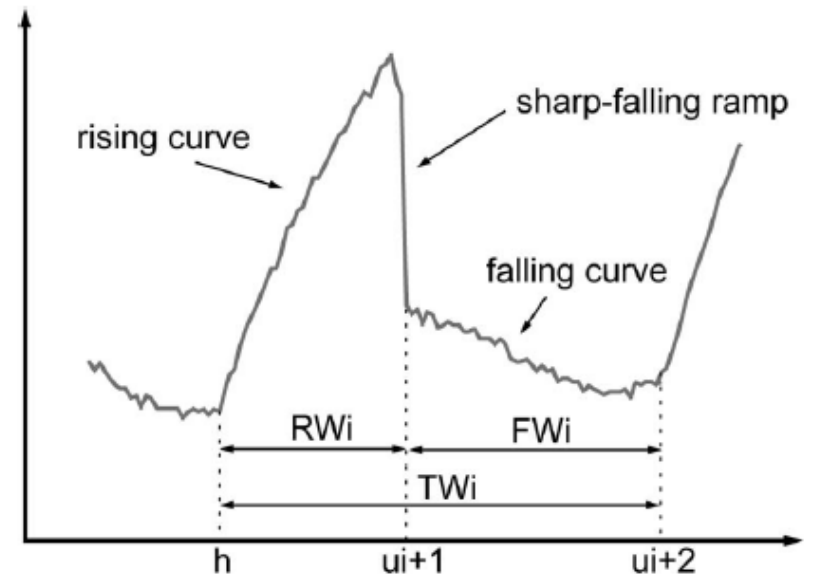

Fig. 2. Enlargement of the current of a brushed dc motor [17].

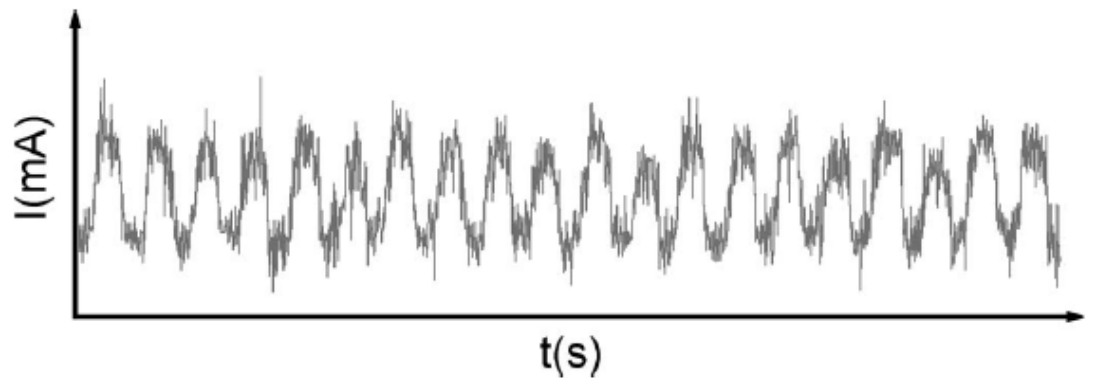

Fig. 3. Enlargement of the real current of a brushed dc motor.

Fig. 2 shows the magnification of the current signal representation of the brushed dc motor current. As this figure illustrates, the ripple component is not a sinusoidal signal but has a rise $RWi$, a steep fall, and another slight fall $FWi$ [17]. Usually, those three sections of the current are not as clear, and sharp drops may not be visible in these situations. Either way, the shape of the ripple component shown in Fig. 2 is ideal, and in this figure, it is assumed that the bandwidth used for the display of the same is sufficiently high. This shape (Fig. 2) is known as an undulation, a pulse, or a commutation. The pulses that appear in the current graph are associated with the brush connection change of the commutator bar in the commutator. In this instant, the coil is short circuited. This action is also related to the instant when the EMF induced in the coil, which is connected to both commutator bars where they produce switching, is zero [18]. Thus, it is possible to measure the movement of a shaft by taking into account the number of pulses over the current signal of the brushed dc motor. In addition, if the time is monitored, the speed can be measured.

Fig. 2 shows an ideal situation without noise, but in practice, the current signal over a brushed dc motor is similar to that in Fig. 3. There are multiple noise types such as that added by the dc motor itself, that associated to the power supply, or those induced by other nearby elements. The noise sources or disturbances sometimes cause the following: 1) false pulses or double pulses in the current and 2) ghost pulses or merging pulses in the current. The false pulses are pulses that appear in the current resulting from noise and do not belong to the ripple component. The ghost pulses are pulses that belong to the ripple component, but the noise masks them in the current [19]. At a low speed, the noise becomes ever more important because the amplitude of the ripple component is smaller, and can be comparable to the noise present in the current. Consequently, to implement an effective system, it is necessary to detect false and ghost pulses. Next, the false pulses must be discarded, and the ghost pulses must be taken into account as a regular pulse ripple. This is necessary in applications where there is great amount of noise, such as noisy industrial applications, or where precision is important, such as robotic applications.

This paper proposes a method based on pattern recognition techniques to detect the pulses of the current of a brushed dc motor and, with this information, to estimate the speed and position of the brushed dc motor. Brushed dc motors are dc motors that have a mechanical commutator. Also, the proposed method is able to detect ghost pulses and to discard false pulses. The classifier support vector machine (SVM) is used for two reasons: First, it has the ability to generalize, and second, it is free of the overfitting problem [20].

In pulse detection systems for estimating speed and position, it is necessary to establish the pulse start time. Some works establish this pulse start time as the instant that the current crosses the mean current value [21]. In the proposed method, the researchers consider the pulse start time to be when the current reaches its maximum value.

Finally, this paper will evaluate the accuracy of the proposed method. Accuracy is measured by comparing the speed and the position estimated by the method with the real speed and real position collected using a high-resolution encoder.

The procedure and tests are described in more details in the following sections. Section II presents the basic theory of SVM for classification. Section III presents the proposed method and algorithm for the training method. Section IV presents a comparison of the proposed method with others used in scientific literature. Section V presents different tests to measure the accuracy and the results obtained. Finally, Section VI presents the conclusions derived from this research.

## II. SVMs for Classification

SVM theory was initially developed by Vapnik [22]. It is a learning machine based on statistical theory and is used for classification and regression. Unlike traditional learning approaches, which are based on empirical risk minimization, SVM is based on structural risk minimization (SRM). The SRM principle improves the generalization ability and avoids the problem of overfitting [20].

Taylor and Cristianini [23] and Abe [24] reviewed the SVM theory. SVM is a linear classifier that establishes a hyperplane to separate two classes. This hyperplane maximizes the margin.

Given a training set or training samples $\{x_i, y_i\}$ with $i = 1, 2, \ldots, M$, where $x_i$ denotes the $d$-dimensional column vectors that are the classifier inputs, $y_i$ denotes the class labels that are the classifier outputs and can be $+1$ or $-1$ to $Class+1$ and $Class-1$, respectively, and $M$ is the size of the training set, the linear function that separates both classes is

$$D(x) = \text{sign}\,(f(x)) \tag{1}$$

$$f(x) = w^{\mathrm{T}} x + b \tag{2}$$

where $w$ is a $d$-dimensional column vector which has the same dimension as the $x$ column vector, $b$ is a scalar, $\text{sign}(\cdot)$ is the sign function, the operator $(\cdot)^{\mathrm{T}}$ is the transpose operator, and $D(\boldsymbol{x})$ returns the label of the class that is classified $\boldsymbol{x}$. The hyperplane that separates both classes is given by $f(\boldsymbol{x}) = 0$. So, vector $\boldsymbol{w}$ and scalar $b$ determine the position of the separating hyperplane, $\boldsymbol{w}$ determines the orientation, and $b$ determines the hyperplane separation to the origin reference. When the training set is linearly separable, (3) must be met, and samples of the training set will be correctly classified

$$y_i f(\boldsymbol{x}_i) \geq 1 \qquad \forall\, i = 1, 2, \ldots, M. \tag{3}$$

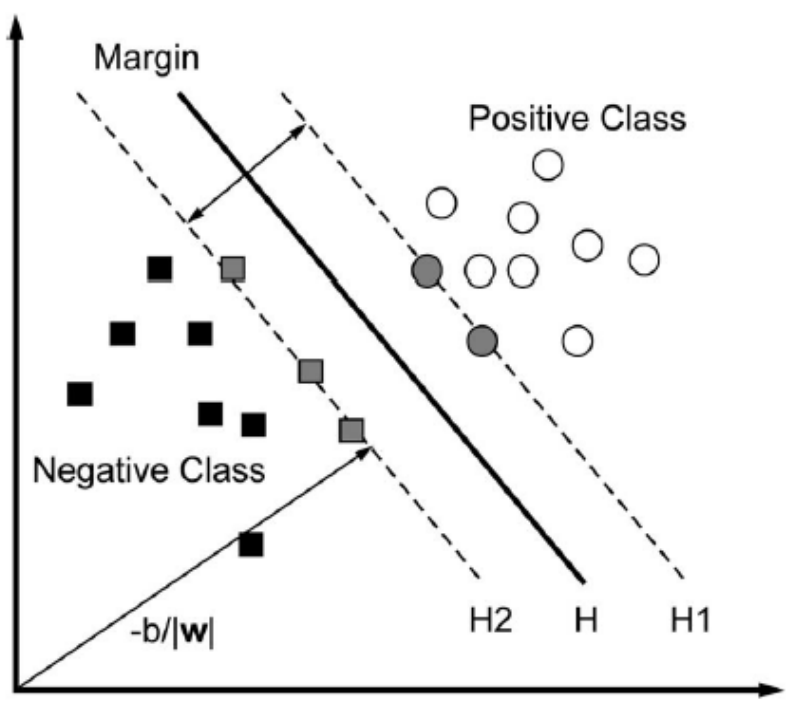

Fig. 4. Separation of two classes by an SVM.

The hyperplane that has the maximum distance between itself and the closest samples is the maximum margin hyperplane, called the optimal separating hyperplane. The classifier with the optimal hyperplane is the classifier with the greatest generalization ability. To find the optimal hyperplane, it is convenient to use three parallel hyperplanes $H$, $H_1$, and $H_2$ (see Fig. 4), such that

$$H : f(\boldsymbol{x}) = \boldsymbol{w}^{\mathrm{T}}\boldsymbol{x} + b = 0 \tag{4}$$

$$H_1 : f(\boldsymbol{x}) = \boldsymbol{w}^{\mathrm{T}}\boldsymbol{x} + b = +1 \tag{5}$$

$$H_2 : f(\boldsymbol{x}) = \boldsymbol{w}^{\mathrm{T}}\boldsymbol{x} + b = -1. \tag{6}$$

The hyperplanes $H_1$ and $H_2$ must contain some training samples. These training samples are known as support vectors and contain all the necessary information to build the classifier hyperplane. The distance between $H_1$ and $H_2$ is called the margin and is $2/\|\boldsymbol{w}\|$. Thus, finding the hyperplane with the maximal margin is equivalent to minimizing the following function:

$$L(\boldsymbol{w}) = \|\boldsymbol{w}\|^2/2 \tag{7}$$

also taking into account the constraint (3). This function can be minimized using the Lagrange multipliers. Then, the problem becomes to minimize (8) with respect to $\boldsymbol{w}$ and $b$ and to maximize (8) with respect to $\boldsymbol{\alpha}$, where $\boldsymbol{\alpha}$ denotes the Lagrange multipliers

$$L(\boldsymbol{w}, b, \boldsymbol{\alpha}) = \|\boldsymbol{w}\|^2/2 - \sum_{i=1}^{M} \alpha_i \left( y_i(\boldsymbol{w}^{\mathrm{T}}\boldsymbol{x}_i + b) - 1 \right). \tag{8}$$

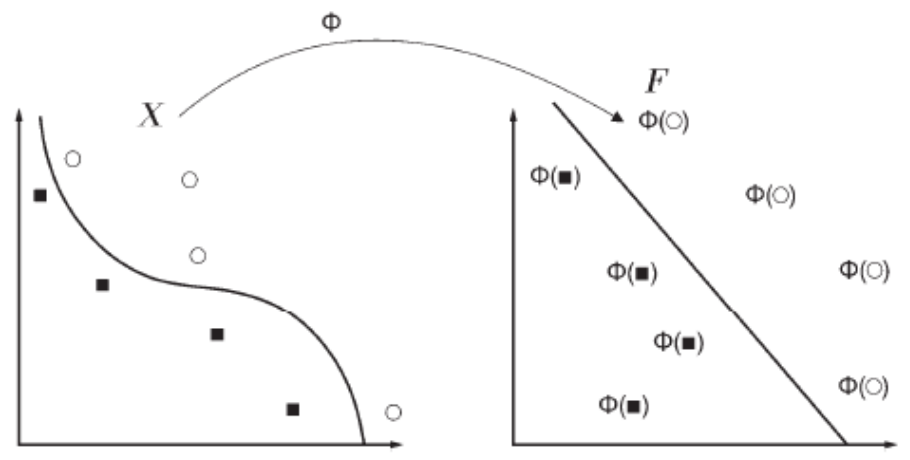

Fig. 5. Nonlinear transformation from input space to high-dimensional feature space.

TABLE I
STANDARD KERNEL FUNCTIONS

| Kernel function | $K(\boldsymbol{x}_i, \boldsymbol{x}_j)$ |
|---|---|
| Linear | $\boldsymbol{x}_i^T \cdot \boldsymbol{x}_j$ |
| Polynomial | $(\boldsymbol{x}_i^T \cdot \boldsymbol{x}_j + 1)^d, d > 0$ |
| Gaussian RBF | $\exp\left(-\|\boldsymbol{x}_i - \boldsymbol{x}_j\|^2/2\sigma^2\right)$ |
| Sigmoid | $\tanh(\gamma \boldsymbol{x}_i^T \cdot \boldsymbol{x}_j + r)$ |

This problem must satisfy the Karush–Kuhn–Tuker (KKT) conditions, which are

$$\partial L(\boldsymbol{w}, b, \boldsymbol{\alpha})/\partial \boldsymbol{w} = \boldsymbol{w} - \sum_{i=1}^{M} \alpha_i y_i \boldsymbol{x}_i = 0 \tag{9}$$

$$\partial L(\boldsymbol{w}, b, \boldsymbol{\alpha})/\partial b = -\sum_{i=1}^{M} \alpha_i y_i = 0 \tag{10}$$

$$\alpha_i \left[ y_i(\boldsymbol{w}^{\mathrm{T}}\boldsymbol{x}_i + b) - 1 \right] = 0 \qquad \forall\, i = 1, 2, \ldots, M \tag{11}$$

$$\alpha_i \geq 0 \qquad \forall\, i = 1, 2, \ldots, M. \tag{12}$$

Substitution of the KKT conditions into (8) gives the dual problem

$$L_{\mathrm{dual}}(\boldsymbol{\alpha}) = \sum_{i=1}^{M} \alpha_i - \frac{1}{2} \sum_{i=1}^{M} \sum_{j=1}^{M} \alpha_i \alpha_j y_i y_j \boldsymbol{x}_i^{\mathrm{T}} \boldsymbol{x}_j \tag{13}$$

with the following restrictions:

$$\sum_{i=1}^{M} \alpha_i y_i = 0 \tag{14}$$

$$\alpha_i \geq 0 \qquad \forall\, i = 1, 2, \ldots, M. \tag{15}$$

Substituting (9) into (2) gives the dual form of the decision function of the following classifier:

$$f(\boldsymbol{x}) = \sum_{i=1}^{M} \alpha_i y_i \boldsymbol{x}_i^{\mathrm{T}} \boldsymbol{x} + b. \tag{16}$$

When classes are not linearly separable, the samples are transformed into a high-dimensional space, where the linear class separation is possible. The data transformation is accomplished via function $\varphi(\cdot)$, which takes the data from the input space to a feature space where classes are linearly separable. An example is shown in Fig. 5. The inner product of two transformed samples is replaced by the kernel function $K(\boldsymbol{x}_i, \boldsymbol{x}_j) = \phi^{\mathrm{T}}(\boldsymbol{x}_i)\phi(\boldsymbol{x}_j)$. Then, the problem given by (13) and (16) becomes

$$L_{\mathrm{dual}}(\boldsymbol{\alpha}) = \sum_{i=1}^{M} \alpha_i - \frac{1}{2}\sum_{i=1}^{M}\sum_{j=1}^{M} \alpha_i\alpha_j y_i y_j K(\boldsymbol{x}_i, \boldsymbol{x}_j) \quad (17)$$

$$f(\boldsymbol{x}) = \sum_{i=1}^{M} \alpha_i y_i K(\boldsymbol{x}_i, \boldsymbol{x}) + b \quad (18)$$

respectively.

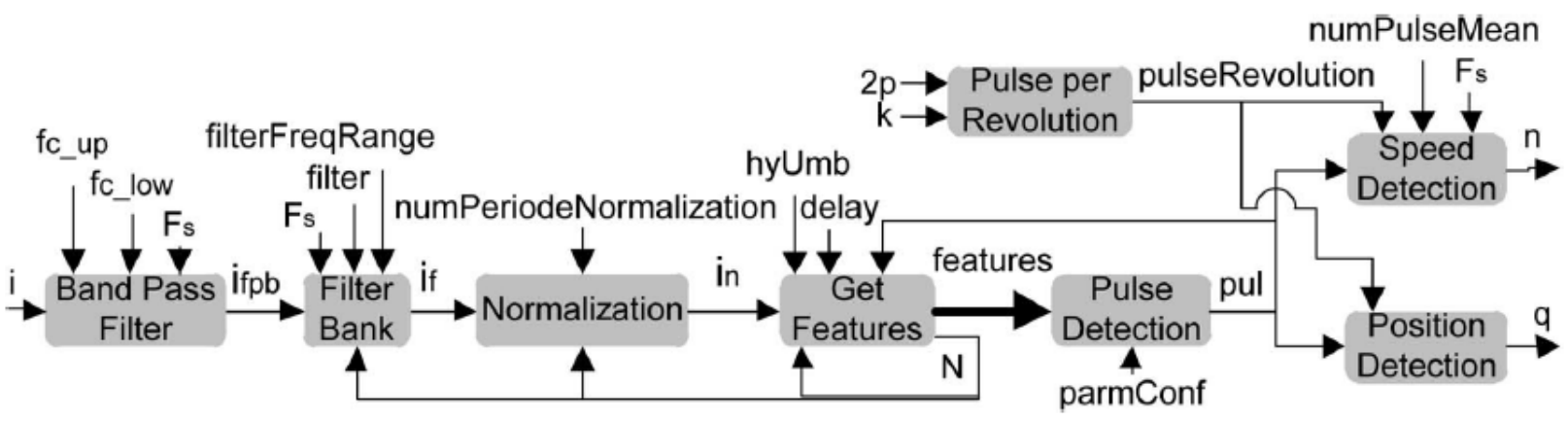

Fig. 6. Block diagram of the proposed method.

In general, the function $\varphi(\cdot)$ is not necessary and may not be known because the kernel function is used instead. The kernel functions represent a valid inner product when they meet the Mercer conditions. The standard kernel functions are shown in Table I.

In previous ideal conditions, it was assumed that classes were separable, and therefore, (3) was always satisfied in the feature space. In real problems, however, it is not always true, and it is necessary to introduce some loss variables $\xi_i$. In this case, (3) becomes

$$y_i f(\boldsymbol{x}_i) \geq 1 - \xi_i \qquad \forall\, i = 1, 2, \ldots, M. \quad (19)$$

The function to optimize in this case is

$$L(\boldsymbol{w}) = \|\boldsymbol{w}\|^2/2 + C\sum_{i=1}^{M} \xi_i \quad (20)$$

where $C$ is a penalty factor. When solving this optimization problem, the same dual problem appears as did in (13), but now subject to the following restrictions:

$$\sum_{i=1}^{M} \alpha_i y_i = 0 \quad (21)$$

$$0 \leq \alpha_i \leq C \qquad \forall\, i = 1, 2, \ldots, M. \quad (22)$$

The resolution of (13) subject to restrictions (21) and (22) is known as SVM classifier training. The training is carried out using samples of the training set in order for $\alpha_i$ and $b$ to solve the dual problem. There are different training algorithms, but the most popular is sequential minimal optimization (SMO) [25].

## III. Proposed Method

The main objective of this paper is to propose a new method for sensorless estimation of speed and position based on the ripple component in brushed dc motors. This method should obtain the pulse of the current, discard false pulses, and detect ghost pulses. It is based on pattern recognition techniques used to detect the pulse of the current with an SVM classifier.

This section explains the proposed method and the procedure to train the method. The method uses an SVM classifier, which needs to be trained.

### A. Method

Fig. 6 shows the block diagram of the proposed method. The vertical arrows in this figure are the parameters, and the horizontal arrows are the processed data. The method processes the digitized samples of the current of a brushed dc motor and consists of the next eight blocks.

*1) Bandpass Filter Block:* This block filters the noise of the brushed dc motor outside the range of possible ripple frequencies. It is composed of a passband filter. The lower cutoff frequency $f_{c_\mathrm{low}}$ is lower than the lowest ripple frequency. The upper cutoff frequency $f_{c_\mathrm{up}}$ is greater than the highest ripple frequency. The parameters of this block are $f_{c_\mathrm{low}}$, $f_{c_\mathrm{up}}$, and $F_s$. The parameter $F_s$ is the sampling frequency of the system, and it is necessary because the filter is digital (see Fig. 6).

In practice, the filter of this block is usually complemented with an antialiasing filter that operates before the A/D converter stage. In this situation, this block can sometimes be eliminated and substituted by this antialiasing filter.

The pulse width modulation (PWM) signal sometimes used to supply brushed dc motors can be considered noise. This noise can be filtered by the *Bandpass Filter Block* if the frequency of the PWM signal does not match the possible frequency ripple. Thus, the frequency of the PWM signal should be set higher than the maximum ripple frequency so that it can be filtered correctly.

*2) Filter Bank Block:* The *Filter Bank* block removes the noise of the current of the brushed dc motor which is within the range of possible ripple frequencies. The noise outside this range is removed by the previous block. This *Filter Bank* block is optional, and it will only be present if the range of possible ripple frequencies has significant noise. The noise in this range can be internal or external, i.e., the noise source may or may not be produced by the brushed dc motor.

The *Filter Bank* block is composed of a filter bank. Only one filter of this filter bank filters the incoming signal, and the filter selection is done according to the ripple frequency detected in the previous iteration. This block has as its input the discrete period of ripple component $N$ that is related to the ripple frequency $f_r$ according to

$$f_r = F_s/N. \tag{23}$$

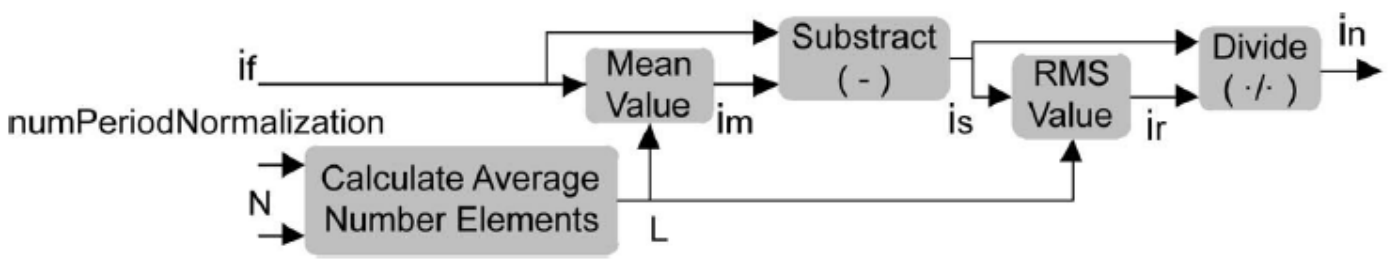

Fig. 7. Block diagram of the *Normalization* block.

The block parameters are $filters$ and $filterFreqRange$. The $filter$ parameter is a list that indicates the number of filters in the filter bank and their lower and upper cutoff frequencies. The $filterFreqRange$ parameter indicates the range in which each filter is used according to the ripple frequency detected in the previous iteration. The configuration of this *Filter Bank* block is given by the brushed dc motor characteristics and the operating environment, i.e., it depends on how much noise the brushed dc motor current has.

*3) Normalization Block:* The dc component and the alternate component of the brushed dc motor current are variable signals. The variations depend on factors such as the load and speed of the brushed dc motor. The *Normalization* block removes this effect and makes the current vary between two fixed values. It also removes the dc component that the current may have after filtration.

The *Normalization* block normalizes the variance of current to $+1$ interval. This block normalizes the variance instead of the amplitude to become more robust in the presence of spurious and impulsive signals. The diagram of this block is shown in Fig. 7. The *Mean Value* block computes the mean value of the last $L$ samples of the current according to

$$i_m[n] = \frac{1}{L[n]} \sum_{k=0}^{L[n]-1} i_f[n-k]. \tag{24}$$

The *Subtract* block removes the dc component of the current according to

$$i_s[n] = i_f[n] - i_m[n]. \tag{25}$$

The *RMS Value* block calculates the variance of the last $L$ samples of the current without the dc component according to

$$i_r[n] = \sqrt{\frac{1}{L[n]} \sum_{k=0}^{L[n]-1} (i_s[n-k])^2}. \tag{26}$$

Finally, the *Divide* block normalizes the variance dividing the current without the dc component by its variance according to

$$i_n[n] = i_m[n]/i_r[n]. \tag{27}$$

The *Calculate Average Number Elements* block calculates the value of $L$ according to

$$L[n] = numPeriodeNormalization \cdot N[n-1]. \tag{28}$$

The parameters of the last block are $N[n-1]$ and $numPeriodeNormalization$. The first parameter is the discrete period of the ripple component detected in the previous iteration. The second parameter is the number of periods used to calculate the mean and variance values and variance or rms value of the current. Larger values of this parameter will make the system more robust against motor noise. However, larger values make a poorer response to the speed variation of the brushed dc motor. In contrast, small values of this parameter will make the system estimate the mean and variance of the current incorrectly. An adequate value for this parameter is between five and ten.

*4) Get Features Block:* The *Get Features* block obtains the most important features of the current that have been filtered and normalized. Later, these features will be used by the *Pulse Detection* block.

Initially, the *Get Features* block delays the current signal according to

$$x[n] = i_n[n - delay] \tag{29}$$

where the delay parameter is the delayed sample number. This is done because it is necessary to have the present and future samples of the signal current in order to obtain some specific features. The features that this block extracts are as follows.

1) **Slope change** ($sC$): This feature measures the variation of the slope's current by means of

$$sC = \frac{\sum_{k=-M}^{M} (x[n] - x[n-k])}{\sum_{k=-M}^{M} |x[n] - x[n-k]|}. \tag{30}$$

   In this, $M$ is one-half of the window's size in which the feature is measured. The denominator is used to normalize the feature value, and it varies between $\pm 1$.

2) **Maximum local** ($m$): This feature determines if the actual current sample is a maximum or not. The value of this feature is obtained from

$$m = \frac{\sum_{k=-M}^{M} higher\,(x[n], x[n-k])}{2M} \tag{31}$$

$$higher(y, z) = \begin{cases} 1, & \text{if } y > z \\ 0, & \text{otherwise.} \end{cases} \tag{32}$$

The denominator of (31) is used to normalize the feature value to the $\pm 1$ range.

3) **Compare with zero** ($cZ$): This feature determines if the normalized current sample, from which the dc component is subtracted, is positive or negative. In order for the noise to not affect the feature measurement, the comparator is implemented with the next hysteresis function

$$cZ[n] = \begin{cases} 1, & \text{if } cZ[n-1] = 1 \ y \ x[n] > -hyUmb \\ 1, & \text{if } cZ[n-1] = 0 \ y \ x[n] > hyUmb \\ 0, & \text{if } cZ[n-1] = 1 \ y \ x[n] < -hyUmb \\ 0, & \text{if } cZ[n-1] = 0 \ y \ x[n] < hyUmb. \end{cases} \quad (33)$$

In this equation, the $hyUmb$ parameter is the hysteresis threshold, and adequate values for this parameter are between 0.3 and 0.5.

4) **Shape similarity** ($s$): This feature determines if the actual current sample is a maximum or not, by means of a correlation between the ideal and real shapes of the brushed dc motor current. The feature value is obtained from

$$s[n] = \frac{\sum_{k=0}^{N-1} \left(x[n-k] x_p \left[k, N[n-1]\right]\right)}{\sqrt{\sum_{k=0}^{N-1} x^2[n] \sum_{k=0}^{N-1} x_p^2 \left[k, N[n-1]\right]}} \quad (34)$$

where $x_p$ is the ideal shape of a maximum obtained from

$$x_p \left[k, N[n-1]\right] = \cos\left(\frac{2\pi k}{N[n-1]}\right). \quad (35)$$

In both equations, $N[n-1]$ is the discrete period of the ripple component detected in the previous iteration.

5) **Rising edge** ($rE$): This feature determines if the current has changed from a negative to a positive value since the last detected pulse. The value of this feature is obtained from

$$rE[n] = \overline{pul[n-1]} \cdot \left(fS[n-1] + cZ[n] \cdot \overline{cZ[n-1]}\right) \quad (36)$$

where $pul[n-1]$ is a logical value that indicates if a pulse was detected in the previous iteration. The operator $\bar{\cdot}$ is the negation operator; the result is zero if the input is different from zero, and it is one if the input is zero.

6) **Falling edge** ($fE$): This feature determines if the current has changed from a positive to a negative value since the last detected pulse. The value of this feature is obtained from

$$fE[n] = \overline{pul[n-1]} \cdot \left(fE[n-1] + cZ[n] \cdot cZ[n-1]\right). \quad (37)$$

7) **Zero crossing distance** ($zCD$): This feature measures the number of normalized samples since the last rising edge. If a rising edge has not yet been detected, then its value is zero. The value of this feature is obtained from

$$zCD[n] = \overline{pul[n-1]} \cdot \left(rE[n]/N[n-1] + zCD[n-1]\right). \quad (38)$$

8) **Length wave time** ($Lwt$): This feature measures the number of normalized samples since the last detected pulse. The value of this feature is obtained from

$$Lwt[n] = \overline{pul[n-1]} \cdot \left(1/N[n-1] + Lwt[n-1]\right). \quad (39)$$

9) **Length wave amplitude** ($Lw\alpha$): This feature measures the distance accumulated by the current amplitude since the last detected pulse. The value of this feature is obtained from

$$Lwa[n] = \overline{pul[n-1]} \cdot \left(|x[n] - x[n-1]| + Lwa[n-1]\right). \quad (40)$$

In this case, the feature is not explicitly normalized because the current was normalized by the *Normalization* block.

Many of these features give redundant information, but because they are obtained in different ways, noise will affect them differently. Therefore, all these features are used, and the *Pulse Detection* block will fuse all information contained in the features to determine if the actual sample corresponds to a pulse of the current.

Some features use the $M$ parameter. This parameter is determined as one-half of the size of the window that is used to extract the feature. In order to eliminate noise as much as possible, this parameter should be as large as possible. This parameter has two restrictions: 1) It must be lower than one-half of the discrete period of the ripple component detected in the previous iteration, because the filter must not eliminate the current pulses, and 2) it must be lower than the delay parameter, because only delay samples will be available as samples in the future. This parameter is given according to

$$M = \min\left(0.4 \cdot N[n-1], delay\right). \quad (41)$$

The *Get Features* block also obtains $N$. This parameter is the number of samples in a period of the ripple component. The $N$ value is calculated as the number of discrete samples between the last two pulses. The equations for calculating $N$ are

$$N[n] = \overline{pul[n-1]} \cdot N[n-1] + pul[n-1] \cdot N_p[n] \quad (42)$$

$$N_p[n] = \overline{pul[n-1]} \cdot \left(N_p[n-1] + 1\right). \quad (43)$$

*5) Pulse Detection Block:* The *Pulse Detection* block determines with each new sample whether a current pulse appears or not. This block also discards false pulses and detects ghost pulses. The inputs of this block are the features obtained in the *Get Features* block, and the output $pul[n]$ is a Boolean value that indicates if a pulse is detected or not. This block is a classifier based on an SVM. The equations to implement the classifier are (1) and (18). In these equations, the parameters are as follows: 1) $\alpha_i, y_i$ from support vector set; 2) the $b$ parameter; 3) the kernel function $K$; and 4) the kernel parameters. All these parameters are obtained in the training method stage, and they are passed to the *Pulse Detector* block using the $featConf$ parameter.

*6) Speed Estimation Block:* The *Speed Estimation* block calculates the speed of the brushed dc motor, with information

obtained from the *Pulse Detection* block. This speed $n$ is calculated only when a new pulse has been detected. The steps of the algorithm to calculate it are as follows.

1) Update the distance or sample number to the last detected pulse $d$ according to

$$d := d + 1 \tag{44}$$

where the operator := indicates that $d$ is a variable of the algorithm, which is updated in each iteration.

2) If a pulse is detected, add the pulse distance to a list of distances between pulses $\tau_k$, and reset to zero the actual distance since the last pulse. The equations that do this are

$$k := k + 1 \tag{45}$$

$$\tau_k := d \tag{46}$$

$$d := 0. \tag{47}$$

Equation (45) updates the number of pulses detected $k$. Equation (46) adds the distance between the last two pulses to the list of distances. Equation (47) resets the distance to the last pulse.

3) Update and output the speed. The present speed is given according to

$$n = \frac{Fs \cdot numPulseMean}{\sum_{i=0}^{numPulseMean-1} \tau_{k-i}} \cdot \frac{60}{pulseRevlolution} \tag{48}$$

where $numPulseMean$ is the number of the last distance between pulses used to calculate the speed and $pulseRevolution$ is the number of pulses produced in the current per revolution of the brushed dc motor shaft.

*7) Position Estimation Block:* The *Position Estimation* block calculates the position $\theta$ of the brushed dc motor. The steps of the algorithm to calculate the position are as follows.

1) Increment the number of pulses $n_{\text{pulse}}$ if a new pulse has been detected. This is

$$n_{\text{pulse}} := n_{\text{pulse}} + 1. \tag{49}$$

2) Calculate the position of the brushed dc motor in radians according to

$$\theta = 2\pi \cdot n_{\text{pulse}}/pulseRevolution. \tag{50}$$

*8) Pulses per Revolution Computation Block:* The *Pulses per Revolution Computation* block determines the number of pulses of the current per revolution of the brushed dc motor shaft. This value is obtained according to

$$pulseRevolution = 2p \cdot k/\eta \tag{51}$$

where $2p$ is the number of poles, $k$ is the number of commutation bars, and $\eta$ is assumed according to

$$\eta = gcd\{2p, k\} \tag{52}$$

where $gcd$ is the greatest common divisor. These parameters depend only on the motor construction, and they do not vary with time [18].

If the number of commutation bars and the number of poles are not known, the parameter $pulsRevolution$ can be determined experimentally. To get this parameter, the following steps should be followed. First, the brushed dc motor has to rotate at a constant speed. Second, the speed is measured by an encoder. Third, when the brushed dc motor is rotating, the $\tau_k$ that is used in (48) is measured. Finally, (48) is solved for the parameter $pulseRevolution$, where $n$ is the speed measured with an encoder.

### B. Training Method

The training method makes the SVM classifier of the proposed method learn from a training set. The classifier is implemented inside the *Pulse Detector* block (Fig. 6). This block decides for each instant of discrete time whether a current pulse has or has not happened. The inputs of the *Pulse Detector* block are the features got from the current signal by the *Get Features* block, and the output is a Boolean value that indicates if a pulse current has happened. The training sets are feature sets extracted by the *Get Features* block when a pulse happens and feature sets extracted when a pulse does not happen.

The objective of the training method is double. First, it gets training sets, and second, it trains the SVM classifier with the above training samples. Thus, the training method gets parameters of the *Pulse Detector* block. The steps to train the method are as follows.

1) **To measure signals of current related to the motor comportment**: Together with the current, the real speed and the position of the brushed dc motor are measured using a conventional sensor such as a high-resolution optical encoder. This is done on different motor conditions, as many constant speeds, many linear variations of the speed, and many speed steps.
2) **To establish a criterion to end the training process**: This criterion indicates when the training method must finish. The criterion should be established on the speed and/or position estimated by the proposed method, never if a pulse is detected or not for determinate features. For example, a criterion can be that the speed or the position does not exceed a maximum error.
3) **To set the parameters of the method**: The method has some specific parameters, and it is necessary to set the values of these parameters by the system designer. However, the values of the *Pulse Detector* block parameters are not established during this step. They are obtained at the end of the training method, and they are established in Step 7).
4) **To empty the list of training samples** $S$: The list of training samples is emptied. This is done in the first iteration and when a parameter of the method is modified.
5) **To get training samples**: The training samples are taken from the measured signals in Step 1). Each training sample is composed by the features taken by the *Get Features* block and by a Boolean value that indicates whether the training samples correspond to a pulse. The designer decides if the features correspond or not to a pulse and facilitates the Boolean value in this stage. A small interval of the measured signals in Step 1) is randomly selected to get the samples when the $S$ list is empty. However, when the $S$ list is not empty, an interval that does not satisfy the end criterion is taken instead of a random interval.

6) **To add samples to the lists of training samples $S$**: The samples obtained in the previous step are added to the list of training samples $S$.
7) **To train the classifier**: The SVM classifier is trained with the list of training samples $S$. The algorithm used in this paper is the SMO, but another algorithm can be used. In this step, the parameter of the *Pulse Detector* block is obtained.
8) **To get the errors of the method**: The proposed method, with parameters obtained in the previous step, is tested with the measured signals in Step 1). The method returns an estimation of the speed and position of the brushed dc motor. The estimation is compared to the real speed and the real position to estimate the error of the method.
9) **To evaluate the error of the method**: The training end criterion is checked. Step 10) is next when the criterion is achieved. If the criterion is not yet achieved, then it is checked if it is convenient to change any parameter. If any parameter must be changed, Step 3) is the next; if not, Step 5) is the next.
10) **To return the parameters of the SVM classifier**: The parameters of the SVM classifier or *Pulse Detector* block are returned. The parameters are $\alpha_i$, $y_i$, $b$, the kernel function, and the kernel parameters.

This algorithm is iterative. Each iteration adds new training samples to train the SVM classifier. The training samples added in each iteration are the samples obtained from Step 1) that measured the signal where the method did not correctly estimate the speed and position.

The kernel function with better practical results in this paper is the polynomial function. The kernel parameter is the degree, which must be below five for optimal results.

## IV. Comparison With Other Methods

The more popular methods to detect the pulses of current in brushed dc motors are those that use a comparator. These methods compare the current in a brushed dc motor with the dc component of the same current signal. Implementations of these methods are as follows: 1) the one proposed by Ma and Weiss [12] that estimates the dc component as the mean value between the maximum and the minimum of the current; 2) that proposed by Iott and Burke [13] that eliminates the dc component with a high-pass filter and compares the result with zero; and 3) the proposal by Micke *et al.* [16] that obtains the dc component with a low-pass filter whose cutoff frequency is sufficiently low.

The disadvantage of these methods is that they cannot detect ghost pulses and cannot discard false pulses, as Fig. 8 shows. In this figure, (a) represents schematically the pulses of the current and its dc component, and (b) represents the output signal of a comparator that has a pulse by each detected pulse.

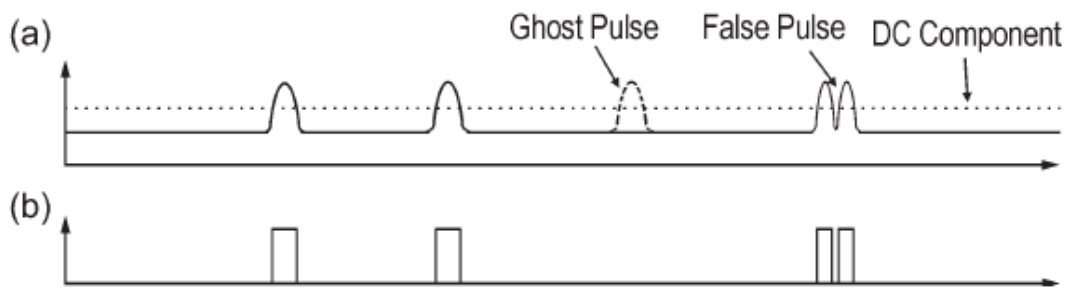

Fig. 8. False and ghost pulses. (a) Pulses of the brushed dc motor current. (b) Pulses detected by a comparator.

A solution to this problem is to combine the pulse detection with a comparator with the distance estimation between pulses through a dynamic model of the brushed dc motor. The dynamic model estimates the speed of a brushed dc motor, which is related to the frequency ripple component [18]. This means that the frequency ripple component is obtained from the estimated speed. The inverse of the frequency is the period, and the period is the temporal distance between two consecutive pulses. A temporal window that must contain the next pulse is estimated using the last temporal position of the pulse and the distance between pulses. The pulses detected before this window starts are discarded because they are estimated as false pulses. A pulse detected during the window is considered valid, and after this, the system starts from the beginning to detect the next pulse. If no pulse is detected during the window, the system determines that a ghost pulse has occurred and a pulse is added. This solution was applied, for example, by Kessler and Schulter [19] and Lutter and Fiedrich [26].

However, the aforementioned solution has the same problems as the sensorless techniques based on the dynamic model of the brushed dc motor. The dynamic model of a brushed dc motor uses some parameters whose values depend on the operating conditions of the brushed dc motor. If these operating conditions vary, these parameters vary in value. If the estimated parameter values are very different from the real parameter values, the method is not able to correctly estimate the distance between pulses or the temporal window. This causes the method to 1) discard some true pulses detected by the comparator and 2) erroneously indicate that a ghost pulse has occurred.

According to the proposed method by Ohishi *et al.* [11], a second solution to the aforementioned problem is the dynamic estimation of dc parameters in the brushed dc motor. However, it leads to a difficult-to-solve nonlinear model that also has a high computational cost. Also, in this solution, the dynamic model of the brushed dc motor needs to monitor the current and the voltage of the brushed dc motor simultaneously. This causes the system to require an additional A/D converter, therefore increasing the system cost.

The proposed method in this paper solves the previous problems with the monitoring of only the current of the brushed dc motor. In this case, the SVM classifier detects current pulses, detects ghost pulses, and discards false pulses. A correct SVM training is essential to allow the system to detect ghost pulses and to discard false pulses. To achieve the objective, the elapsed time since the last detected pulse is introduced in the obtained features. Real examples of the detection of a ghost pulse and the discarding of a false pulse are shown in Figs. 9 and 10, respectively. These figures also show the brushed dc motor current whose variance is normalized to $\pm 1$, the pulses detected

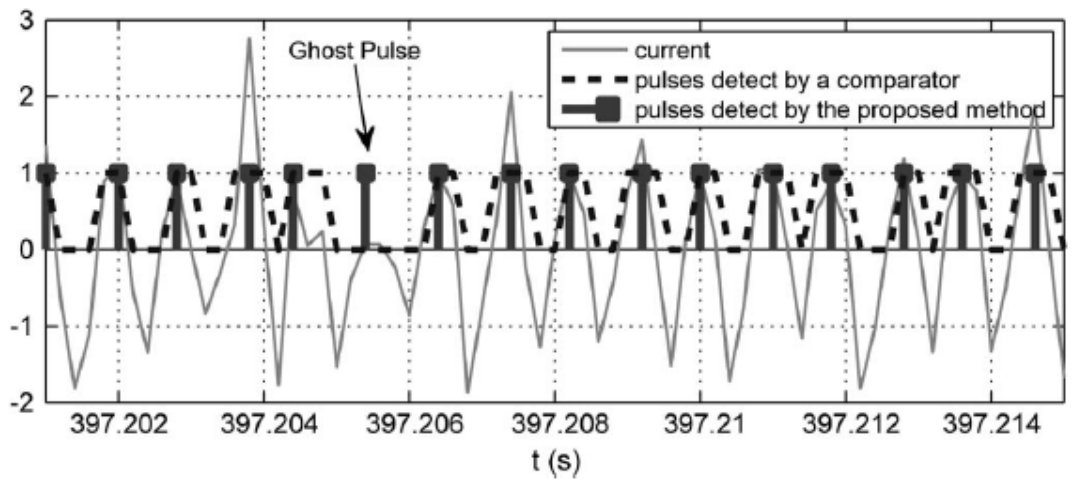

Fig. 9. Real example of the detection of a ghost pulse using the proposed method. The method that uses a comparator does not detect the ghost pulse. In this example, the speed of the brushed dc motor is high.

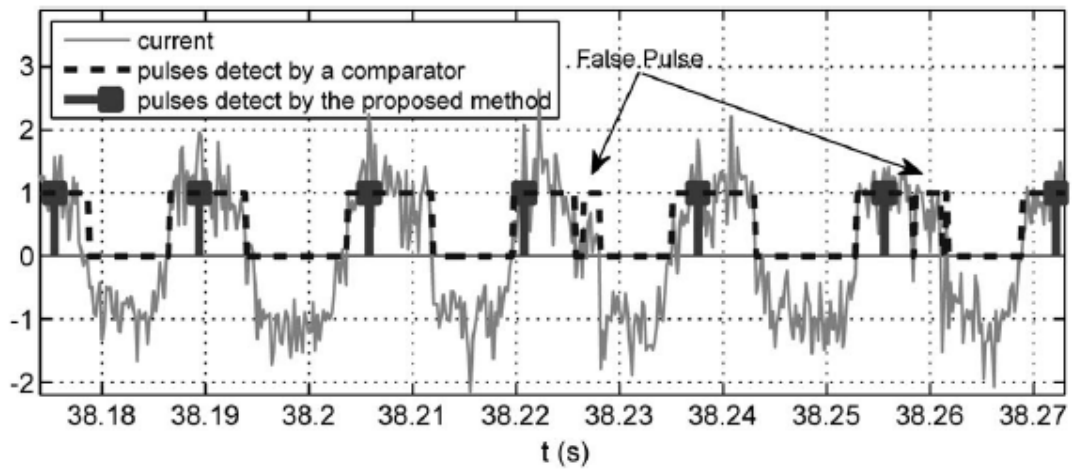

Fig. 10. Real example of the detection of a false pulse using the proposed method. The method that uses a comparator does not discard the false pulse. In this example, the speed of the brushed dc motor is low.

by the proposed method, and the pulses detected by a method that uses a comparator.

Modern microcontrollers or DSP devices can support the computational cost of the proposed method if the method is parallelized. In addition, there are some techniques that decrease the number of operations that need to be realized by the SVM [23], [24]. Also, in order to decrease the hardware cost, the sensorless processing hardware can be shared by several brushed dc motors or by other devices.

## V. Experimental Results

This section presents the procedure for measuring the accuracy of the proposed method and presents the results obtained from the two brushed dc motors that were tested.

### A. Experiment

The accuracy of the proposed methods is measured by comparing the estimated speed and position with the real speed and position of a brushed dc motor in different situations of work: 1) a constant speed; 2) a linear variation of the speed or a constant acceleration; and 3) a speed step.

Fig. 11 shows the architecture hardware. The architecture consists of a brushed dc motor, a current sensor, a data acquisition card, and a PC.

Tests were done with two brushed dc motors, the EMG30 and the 719RE385, whose characteristics are shown in Table II. The sensor of the current was a small resister with only 20 mΩ, usually called a shunt resistor. A low-cost data acquisition card (National Instruments USB-6008) was used to measure and digitalize the signal of the current by means of one of its four differential analog inputs. The sampling frequency was 5 kHz. The data acquisition card was connected to a PC. This processed the signal of the current in order to estimate the speed and the position of the brushed dc motor according to the proposed method. The PC was a laptop with a T8300 microprocessor, 3 GB of RAM, and 320 GB of hard disk space. The operating system was Windows 7, and the development environment was Matlab R2009a. The PC was configured to work in soft real time, and the data acquisition card was configured to acquire the current in a continuous mode. In order to estimate the real speed and position of the brushed dc motor and to compare it with the proposed method, a high-resolution incremental encoder attached to the motor shaft was used. The encoder was connected to the 32-bits counter of the data acquisition card. Fig. 11 does not show the encoder and its connections.

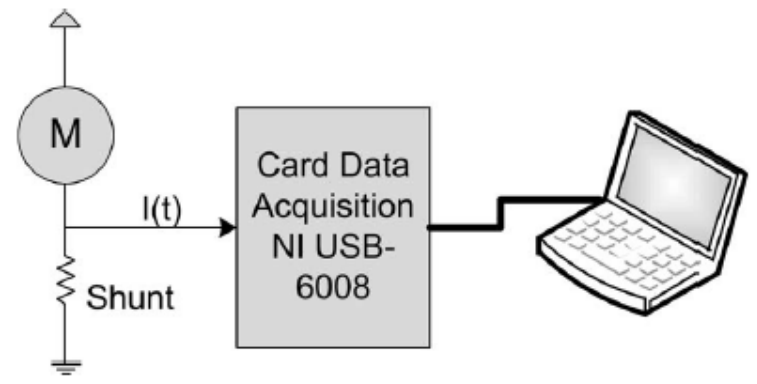

Fig. 11. Architecture hardware.

TABLE II
MOTOR SPECIFICATIONS

| Parameter / Motor | EMG30 | 719RE385 |
|---|---|---|
| Manufacturer | Devantech Ltd. | Como Drills |
| Nominal Voltage | 12 V | 12 V |
| No Load Current | 530 mA | 250 mA |
| Nominal Speed | 6000 r.p.m. | 5000 r.p.m. |
| Poles ($2p$) | 2 | 2 |
| Commutator Bars ($k$) | 3 | 5 |
| Resistance ($R_a$) | 1.8 Ω | 1.5 Ω |
| EMF Constant ($c$) | 0.0178 V/r.p.m. | 0.00101 V/r.p.m. |

### B. Results

The results obtained in the tests are shown separately for the three different behaviors of the brushed dc motors.

*1) Constant Speed:* In this situation, the two brushed dc motors are turned at different constant speeds. In this test, the average error and deviation error of speed, with respect to real speed and position, were obtained. The deviation error is shown in absolute value (in revolutions per minute) and in relative value (in percent). Also, shown are the estimated positions and real positions for several constant speeds.

Tables III and IV show the results obtained in the estimation of the brushed dc motor speed for the EMG30 and 719RE385 dc motors. The average error of the estimate speed is low and constant. However, the absolute error variance increases as the speed increases. However, the relative error variance is always constant. This indicates that absolute error variance is proportional to the speed. Figs. 12 and 13 show the estimated and real positions for the EMG30 and 719RE385 dc motors. For the EMG30 dc motor, the position is shown at the constant

TABLE III
SPEED MEASUREMENT ERRORS WITH THE EMG30 DC MOTOR

| Real Speed (r.p.m.) | Average error | | Deviation | |
|---|---|---|---|---|
| | Absolute (r.p.m.) | Relative (%) | Absolute (r.p.m.) | Relative (%) |
| 501 | 1.03 | 0.21 | 3.51 | 0.70 |
| 807 | 1.09 | 0.14 | 5.86 | 0.73 |
| 1044 | 0.88 | 0.08 | 6.72 | 0.64 |
| 1526 | 1.79 | 0.12 | 8.62 | 0.56 |
| 2028 | 0.28 | 0.01 | 12.22 | 0.60 |
| 3082 | 5.72 | 0.19 | 7.22 | 0.23 |
| 4051 | 1.01 | 0.02 | 20.45 | 0.50 |
| 5055 | 3.26 | 0.06 | 21.51 | 0.43 |
| 6088 | 2.39 | 0.04 | 19.11 | 0.31 |
| 8041 | 4.01 | 0.05 | 32.14 | 0.40 |
| 9017 | 3.96 | 0.04 | 34.62 | 0.38 |
| 10117 | 1.96 | 0.02 | 36.13 | 0.36 |
| 11097 | 5.79 | 0.05 | 29.74 | 0.27 |

TABLE IV
SPEED MEASUREMENT ERRORS WITH THE 719RE385 DC MOTOR

| Real Speed (r.p.m.) | Average error | | Deviation | |
|---|---|---|---|---|
| | Absolute (r.p.m.) | Relative (%) | Absolute (r.p.m.) | Relative (%) |
| 592 | 0.18 | 0.03 | 2.41 | 0.41 |
| 858 | 0.36 | 0.04 | 2.52 | 0.29 |
| 1029 | 0.43 | 0.04 | 2.45 | 0.24 |
| 1499 | 0.40 | 0.03 | 3.36 | 0.22 |
| 1971 | 0.74 | 0.04 | 4.73 | 0.24 |
| 3010 | 0.07 | 0.002 | 7.50 | 0.25 |
| 3949 | 2.86 | 0.07 | 10.82 | 0.27 |
| 5012 | 7.51 | 0.15 | 21.25 | 0.42 |
| 6084 | 5.64 | 0.09 | 28.82 | 0.47 |
| 7062 | 12.18 | 0.17 | 36.96 | 0.52 |
| 8017 | 11.49 | 0.14 | 49.48 | 0.62 |
| 8964 | 11.12 | 0.12 | 49.71 | 0.55 |
| 9994 | 17.54 | 0.18 | 76.67 | 0.77 |

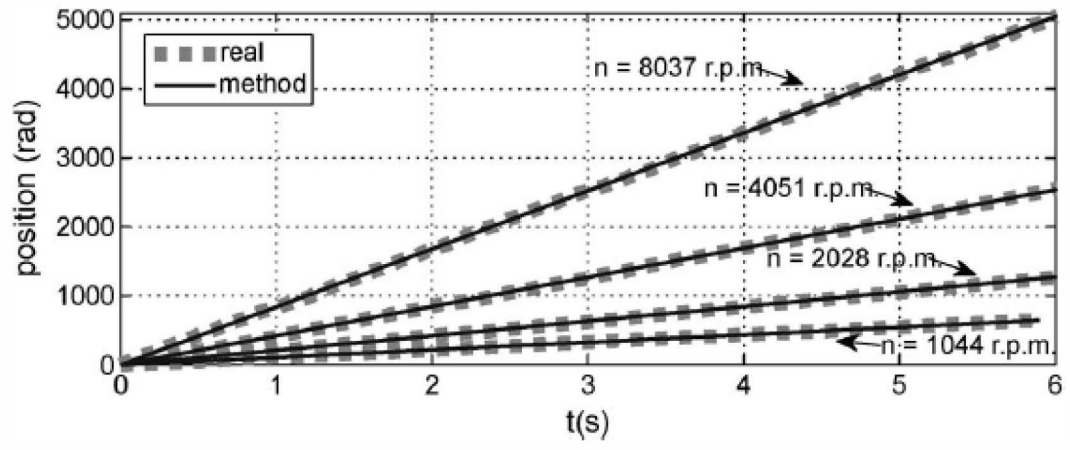

Fig. 12. Position at different constant speeds in the EMG30 dc motor.

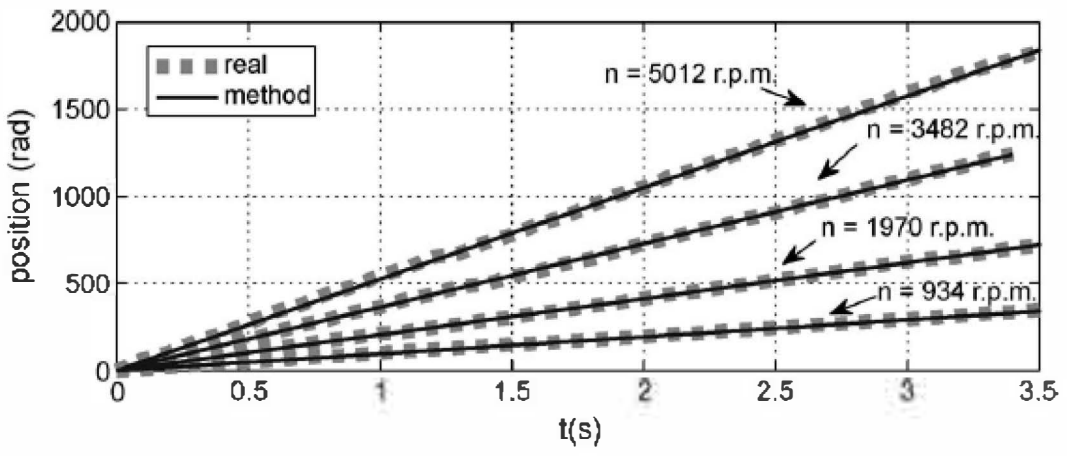

Fig. 13. Position at different constant speeds in the 719RE385 dc motor.

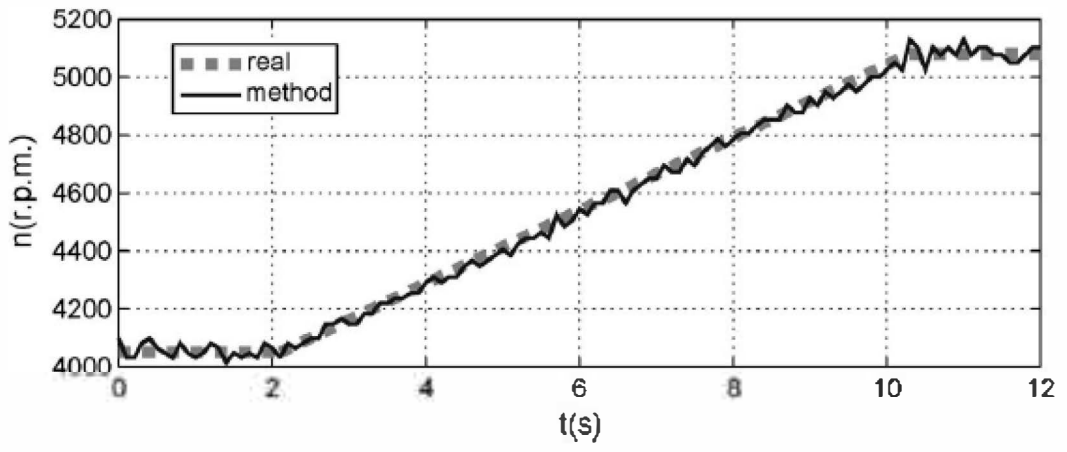

Fig. 14. Speed with linear variation of speed in the EMG30 dc motor.

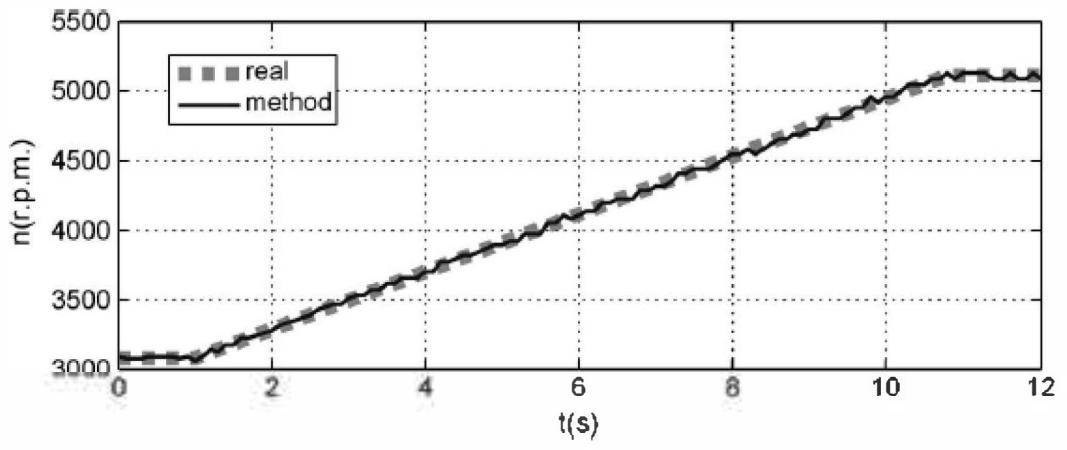

Fig. 15. Speed with linear variation of speed in the 719RE385 dc motor.

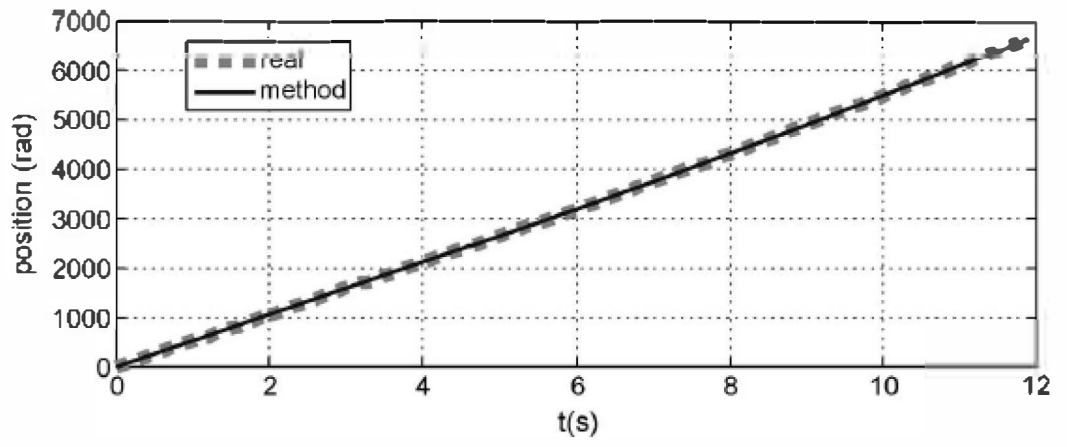

Fig. 16. Position with linear variation of speed in the EMG30 dc motor.

speeds 1044, 2028, 4051, and 8037 r/min, and the average errors of position were 1.81, 7.49, 3.35, and 2.52 rad, respectively. For the 719RE385 dc motor, the positions at the constant speeds 934, 1970, 3482, and 5012 r/min are shown. The average position errors were 3.19, 1.44, 0.47, and 6.50 rad, respectively. These data indicate that the proposed method works correctly in a wide range of speeds.

*2) Constant Acceleration:* In this situation, the speed of the brushed dc motor varies linearly or with a constant acceleration. The results shown are the estimated speeds and positions with respect to the real speeds and positions.

For the EMG30 dc motor, the speed varies linearly from 4100 to 5100 r/min, and for the 719RE385 dc motor, it varies linearly from 3100 to 5100 r/min. Figs. 14 and 15 show the speed results for each dc motor. For the EMG30 dc motor, the average error of the speed was 0.49 r/min, and the variance was 20.03 r/min. For the 719RE385 dc motor, the average error of the speed was 0.16 r/min, and the error variance was 10.57 r/min. Figs. 16 and 17 show the position results for the two dc motors. The average error of position was 2.09 rad for the EMG30 dc motor, and it was 18.98 for the 719RE385 dc motor. All these data indicate that the error is low and that the proposed method correctly estimates the speed and position of a brushed dc motor when the speed varies slowly.

*3) Speed Step:* In the third work situation, a speed step is produced in the speed of the brushed dc motor. The results

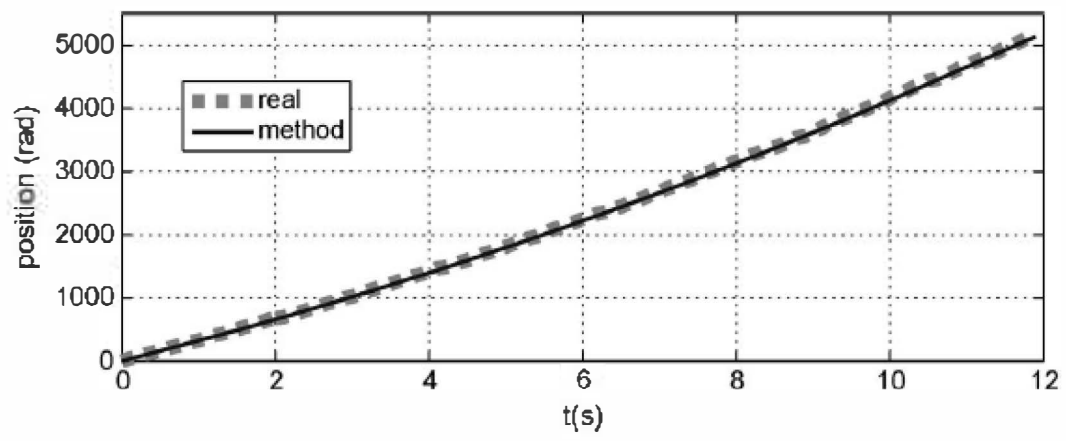

Fig. 17. Position with linear variation of speed in the 719RE385 dc motor.

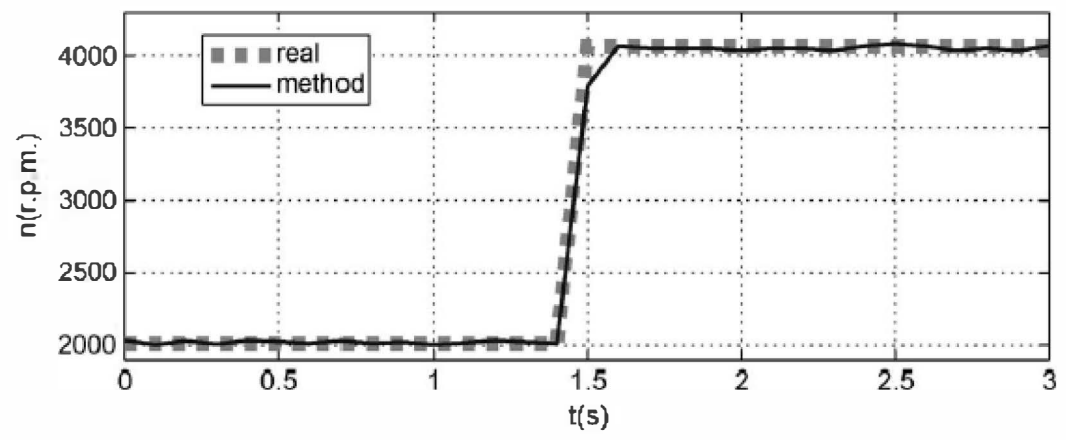

Fig. 18. Speed with speed step in the EMG30 dc motor.

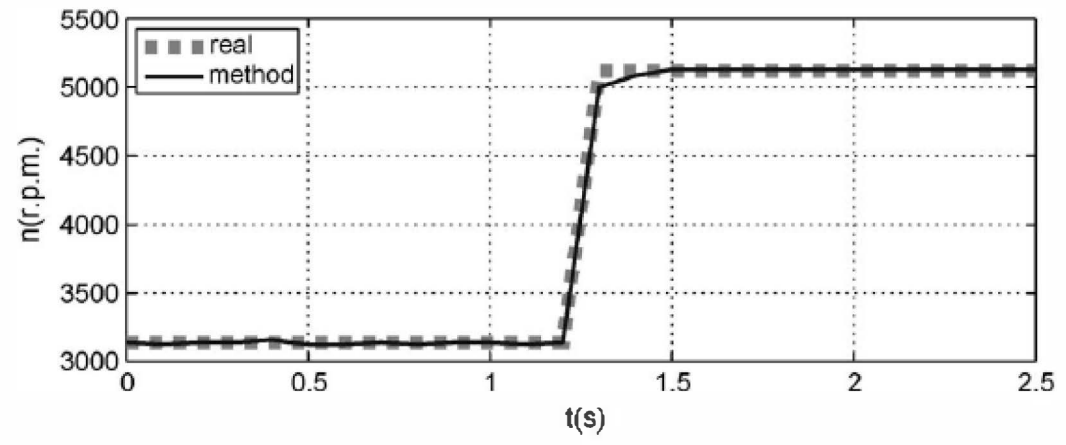

Fig. 19. Speed with speed step in the 719RE385 dc motor.

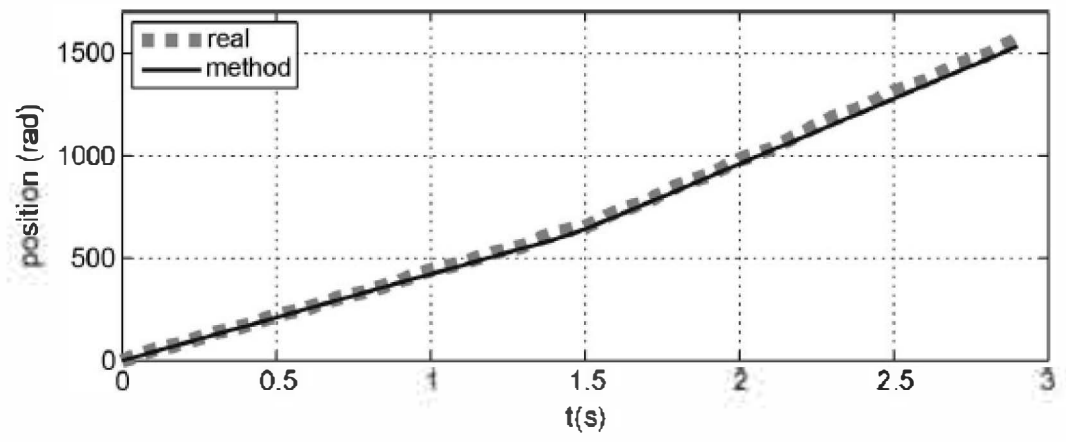

Fig. 20. Position with speed step in the EMG30 dc motor.

shown are the estimated speed and position with respect to the real speed and position.

For the EMG30 dc motor, the step speed was from 2000 to 4000 r/min, and for the 719RE385 dc motor, it was from 3100 to 5100 r/min. Figs. 18 and 19 show the estimated and real speeds for the two brushed dc motors. The proposed method took 0.1 s to reach the final value of the speed for the EMG30 dc motor and 0.2 s for the 719RE385 dc motor. Figs. 20 and 21 show the position results for the two dc motors. The average error of position was 15.90 for the EMG30 dc motor, and it was 1.02 for the 719RE385 dc motor. These results indicate that the error is low and that the proposed method is able to estimate correctly the speed and position.

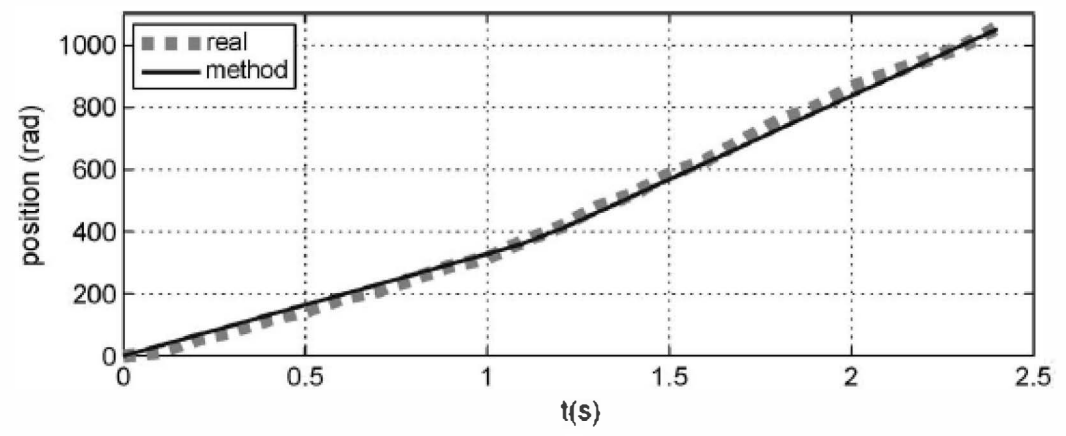

Fig. 21. Position with speed step in the 719RE385 dc motor.

## VI. Conclusion

This paper has presented a new sensorless method for estimating the speed and position of brushed dc motors using SVMs. The method uses sensorless techniques based on the ripple component. The method employs pattern recognition techniques to detect the pulse in brushed dc motor current signals, and it uses SVMs to classify the pulse. To identify the pulses, the method filters, normalizes, and obtains the more important features of the current in a brushed dc motor. Then, the SVM decides in each instance if a pulse has been produced. Finally, the method counts the detected pulses in order to estimate the position and takes the inverse temporal distance between pulses in order to estimate the speed.

This method has an advantage over other existing methods: the ability to detect ghost pulses and to discard false pulses. This is achieved by introducing the time that has elapsed since the last detected pulse into the feature set and by using an SVM as a classifier to detect the pulses.

The experimental results, obtained to validate the proposed method, show that the method works in a wide range of speeds and in different operating conditions, such as linear speed variation and abrupt jumps of speed in a brushed dc motor.

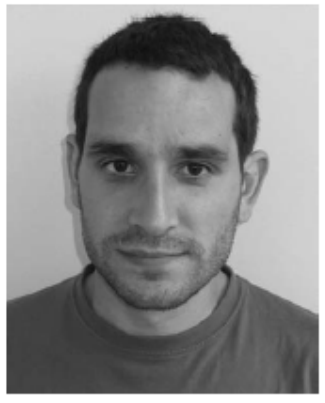

**Ernesto Vázquez-Sánchez** was born in Plasencia, Spain, in 1985. He received the M.S. degree in telecommunication engineering and the M.S. degree in electronic engineering from the University of Valladolid, Valladolid, Spain, in 2008 and 2010, respectively.

Since 2009, he has been as a Researcher with the Department of Signal Theory, Communications and Telematics Engineering, University of Valladolid. His current research interests are communications and sensorless technology for motors.

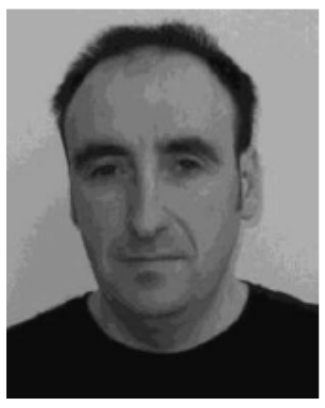

**Jaime Gómez-Gil** was born in Aguilar de Bureba, Spain, in 1971. He received the M.S. degree in telecommunication engineering and the Ph.D. degree in signal theory, communications and telematics engineering from the University of Valladolid, Valladolid, Spain, in 2001 and 2005, respectively.

From 2001 to 2002, he was with the Center for the Development of Telecommunication in Castilla y León (CEDETEL), Valladolid. Since 2002, he has been a full-time Professor with the Department of Signal Theory, Communications and Telematics Engineering, University of Valladolid. His current research is centered on GPS and machine vision applied to agricultural engineering. His other interest areas are sensorless technology and human–computer interface technology.

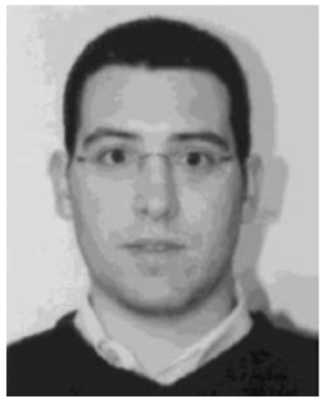

**José Carlos Gamazo-Real** was born in Valladolid, Spain, in 1979. He received the B.S. and two M.S. degrees in electronics and telecommunications engineering from the University of Valladolid, Valladolid, in 2001, 2004, and 2007, respectively.

From 2003 to 2004, he was with the Robotics, Artificial Vision and Real-time Systems Department, Cartif Technology Centre, Valladolid, as an Associate Researcher. From 2006 to 2009, he was with the R&D and Electronics Department, Enerman, S.A., as the Chief Coordinator and the Principal Research Engineer. Since 2009, he has been with the Test Facilities Department, EADS-CASA Cassidian, Madrid, Spain, where he is a Design and Integration Engineer of Avionics Military Air Systems. His current research interests are sensorless control of brushless machines using power electronics and their applications for flight control systems of aircraft aerodynamic surfaces.

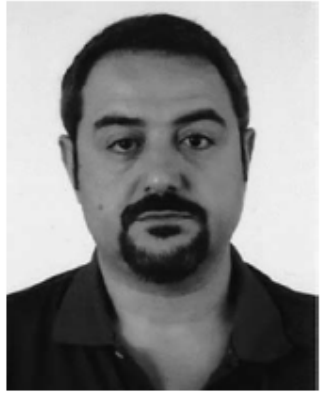

**José Fernando Díez-Higuera** was born in Bercero, Spain, in 1963. He received the M.S. degree in industrial engineering and Ph.D. degree in telecommunications engineering from the University of Valladolid, Valladolid, Spain, in 1991 and 1996, respectively.

From 1991 to 1999, he was an Assistant Professor with the Higher School of Telecommunications Engineering, University of Valladolid, where he has been a Professor of computer programming since 1999 and a Researcher with the Industrial Telematics Group, Department of Signal Theory, Communications and Telematics Engineering, since 1996. His main research interests are biologically inspired visual models, development, business intelligence, intelligent transport systems, and games-based computer science learning.